# CYBER RISK ASSESSMENT FOR CYBER-PHYSICAL SYSTEMS: A REVIEW OF METHODOLOGIES AND RECOMMENDATIONS FOR IMPROVED ASSESSMENT EFFECTIVENESS

Asila AlHarmali[1], Saqib Ali[1], Waqas Aman[1], Omar Hussain[2]

[1]Department of Information Systems, Sultan Qaboos University, Muscat, Oman
[2] School of Business, University of New South Wales, Canberra, Australia

## ABSTRACT

*Cyber-Physical Systems (CPS) integrate physical and embedded systems with information and communication technology systems, monitoring and controlling physical processes with minimal human intervention. The connection to information and communication technology exposes CPS to cyber risks. It is crucial to assess these risks to manage them effectively. This paper reviews scholarly contributions to cyber risk assessment for CPS, analyzing how the assessment approaches were evaluated and investigating to what extent they meet the requirements of effective risk assessment. We identify gaps limiting the effectiveness of the assessment and recommend real-time learning from cybersecurity incidents. Our review covers twenty-eight papers published between 2014 and 2023, selected based on a three-step search. Our findings show that the reviewed cyber risk assessment methodologies revealed limited effectiveness due to multiple factors. These findings provide a foundation for further research to explore and address other factors impacting the quality of cyber risk assessment in CPS.*

## 1. INTRODUCTION

Since the late 18th century, humankind has been continuously improving the capabilities of the physical world from mechanization to mass production to digitization and, recently, to the fourth industrial revolution with Cyber-Physical Systems (CPS) and the Internet of Things and Services. CPS integrates cyber and physical processes, creating new capabilities for the physical environment, such as live communication and process monitoring and control. On the other hand, the emergence of computing and communication components brought new security challenges and exposed CPS physical resources to a range of cyber-attacks [1]. Securing CPS against attacks is vital to avoid catastrophic consequences of critical operations being manipulated or halted by attackers [2]. However, securing information and communication technology systems (ICT) in the context of CPS is a complex task [3]. This complexity stems mostly from the nature of CPS applications, which are safety- and mission-critical, with stringent requirements for high availability and real-time responsiveness [4]. This uptime requirement makes the system sensitive to security updates, reboots, and any security controls' restrictions. Another challenge is the complexity of cyber-physical interactions in a highly interconnected CPS infrastructure, which increases the number of attack points that must be protected [5].

Various security solutions exist for standard IT: the information technology systems consisting of computers and communication networks to process, store, and exchange electronic data. However, these solutions do not apply to CPS system requirements [1]. Various studies have been conducted on developing CPS-specific technical security controls, such as intrusion detection/prevention systems and security information and event management systems [6,7]. Much scholarly work exists on leveraging artificial intelligence (AI) to detect and prevent malicious events and handle monotonous security tasks in a fluctuating CPS environment [8–10]. Despite the capabilities of the existing technical solutions, cybersecurity breaches in CPS are on the rise and show no signs of abating [11–13]. Cyber-attacks have grown in sophistication and complexity, exploiting zero-day vulnerabilities and evading detection [14]. As noticed in [11], cyber-attacks are faced in various CPS sectors, making every OT system a possible target. The motivation of these attackers extends beyond financial gain, with many aiming to harm the public.

Technical security controls alone are insufficient to address the given state of CPS security. A common motto in the security field says that security is 30 percent by technology and 70 percent by management [15]. Therefore, risk management of inevitable cyber-attacks is vital to the organization's cybersecurity program [16]. Risk Assessment is a core function in the risk management process, encompassing the identification, analysis, and evaluation of cyber-attack risks to an organizational operation, assets, individuals, and other organizations [17]. Security risk assessment explains the system security posture and informs related risk management activities and decisions. The more effective an organization can assess the cyber risks, the more rational, effective, and valuable its cybersecurity investments and approach will become [18].

The effectiveness of risk assessment is multifaceted. For instance, effective risk assessment is, in part, determined by validity, how successfully an assessment measures what it is supposed to be measuring [19]. More precisely, valid risk assessment is determined by: (1) how accurate the risk assessment outcomes are compared to the underlying true risk, and (2) the degree to which risk assessment handles uncertainties [20]. Handling uncertainties requires compiling and presenting the missing information, subjective determinations, and assumptions made in the risk assessment process in a way that enables making informed risk management decisions [17]. According to the NIST guide for conducting risk assessment [17], effective risk assessment is also characterized by reliability, the consistency of the risk assessment method, where it can yield the same results when repeated. Reliable risk assessment should be repeatable (repeating the assessment in a manner consistent with prior assessments) and reproducible (producing the same results from the same data across different experts).

Although various cyber risk assessment approaches for CPS have been introduced in previous work, it is unclear how effective they are in assessing cyber risks. A number of available literature review studies present the state-of-the-art cyber risk assessment for CPS, but a detailed evaluation of the effectiveness of the existing approaches was not their primary focus. For example, the systematic literature review study [16] outlines contributions to security risk management in IoT/CPS, including the approaches for the risk assessment step. This study identifies some limitations in the risk identification and analysis techniques but does not analyze them in the context of effective risk assessment requirements.

This paper reviews twenty-eight cyber risk assessment approaches for CPS published in reputable databases between 2014 and 2023. This study seeks to identify the methods the researchers followed to evaluate their proposed cyber risk assessment methodologies and assess the potential effectiveness of the studied risk assessment methods in assessing cyber risks. It also identifies gaps causing limited assessment effectiveness and provides a recommendation of real-time learning from cybersecurity incidents to overcome the most challenging limitations. The rest of this paper is as follows: Section 2 provides essential contextual information on cyber risk assessment for CPS and real-time learning from cybersecurity incidents. We then present the review methodology in section 3. Section 4 provides an analysis of the reviewed risk assessment approaches and answers to the research questions. In Section 5, recommendations to improve cyber risk assessment for CPS are provided. Finally, we conclude the study in section 6.

## 2. BACKGROUND

### 2.1. Cyber Risks in Cyber-Physical Systems

Cyber-physical systems represent a paradigm in automation technology, integrating physical and embedded systems with communication and information technology systems [21]. The integration between operational technology (OT) and IT systems creates intelligent systems that can sense, monitor, and control physical processes at anytime from anywhere. However, this integration exposes the OT infrastructures to cyber risks. Cyber risk is the likelihood of a threat actor, cyber attacker, exploiting a vulnerability within a digital system or network and the resulting impact of successful exploitation. Cyber risks relate to the loss of confidentiality, integrity, or availability of information, data, or information (or control) systems and the potential impacts on organizational operations and assets, individuals, other organizations, and the nation [22]. Cyber risks in CPS may include unauthorized access to critical infrastructure, manipulation of control systems, disruption of critical operations, and compromise of sensitive data. The consequences of cyber risks in CPS, if materialized, extend beyond financial gain to encompass the physical world. Cyber risks in the CPS environment can cause catastrophic impacts, such as manipulating or halting critical physical processes, consequentially harming public health and safety [23].

### 2.2. Risk Assessment

Proactive cyber risk management is vital to the organization's cybersecurity program to prevent the probable risks or reduce their impact to a minimum. Cyber risk assessment is a foundation for informing cybersecurity risk management decisions on preventing cyber risks, mitigating their impact, or effectively handling them when they occur. Risk assessment explains the nature and severity of potential cyber-attacks by answering the question, "What can go wrong?", "What is the likelihood that it would go wrong?", and "What are the consequences if it goes wrong?" [24].

Organizations define a risk assessment methodology that includes a risk assessment process, a risk model, an assessment approach, and an analysis approach [17]. The risk assessment process comprises four steps: preparing for the risk assessment, conducting the assessment, communicating the assessment results, and maintaining the risk assessment over time. The risk assessment is conducted in three tasks: risk identification, analysis, and evaluation. Risk identification is identifying threat sources that could exploit the system vulnerabilities. Risk analysis determines or computes the extent to which an identified threat could harm the system and the likelihood that such events and harm will occur. Risk evaluation rates the system's exposure to risk against the organizational risk tolerance to prioritize risk response.

Due to the highly interconnected CPS infrastructure with complex cyber-physical interactions, the cybersecurity risk assessment methods for general IT systems do not fit the context of CPS [25]. These conventional methods are predominantly static designed to evaluate risks over a set period. CPS requires a cyber risk assessment method capable of assessing the risk at any specific time. A real-time cyber risk assessment that can autonomously learn about the potential cyber risks and adjust to the changing threat landscape and the dynamic nature of the system is crucial for CPS.

### 2.3. Real-Time Learning from Cybersecurity Incidents

Alongside the proactive security risk management process, organizations employ incident response (IR) to take immediate action to a successful attack to minimize effects and expedite system recovery. IR activities include preparing to handle the incidents, detecting signs of incidents and analyzing them, containing the incident and preventing it from spreading, eradicating the root cause, recovering the system operations, and performing post-incident analyses [26]. In cybersecurity, post-incident analysis leverages the organization's experiences with cyber incidents to guide the organization's cybersecurity management process and future cyber incident response activities.

As for the role of IR activities in the security risk assessment process, Shedden [27] says, "Incident response can be a source of concrete data, reflecting what is happening in the organization. This data can be used to inform the risk assessment process, resulting in much more accurate risk assessments and subsequent strategies". Ahmad [28] pointed out the shortcoming of not leveraging opportunities for wider learning, such as improving security risk assessment and security policy development, in existing incident response methodologies. This study highlighted that not drawing broad security lessons can result in little prospect of improving the security of information systems. NIST's guide to OT security proposed leveraging post-incident analysis to update risk assessment with the impact level of the experienced incidents [29].

Real-time analysis of cybersecurity incidents allows learning from the incident as it occurs and avoids erosion of knowledge, which can update the cyber risk assessment with up-to-date, representative, and precise information about cyber risks. Machine Learning (ML), a form of artificial intelligence in which computers gain insights from data to make predictions or decisions, can enable real-time learning from cybersecurity incidents to update the risk assessment process. ML is widely used to develop cybersecurity solutions, such as detecting malicious events and preventing attacks before they commence [30]. ML can automatically learn from different cyber threat information sources, adjust to the changing threat landscape, and identify patterns that may not be immediately obvious to human analysts [30,31].

## 3. RESEARCH METHODS

This study adopted a literature review method to better understand the effectiveness of the existing cyber risk assessment approaches for CPS. The review process started with determining the research question and defining the search string and sources. Inclusion and exclusion criteria were then applied to select twenty-eight relevant studies. The research questions were then answered, and the results were synthesized.

### 3.1. Research Questions

This study primarily seeks to investigate the effectiveness of the existing cyber risk assessment approaches for CPS. Two questions are formulated to achieve this objective. Each question addresses a key aspect of the cyber risk assessment methodologies under study. The first question looks for answers on the methods the researchers followed to evaluate their proposed cyber risk assessment methodologies, the evaluation criteria, and the thoroughness of the evaluation process. The second question investigates the extent to which the studied risk assessment methodologies meet the requirements of effective risk assessment. The effectiveness is evaluated by looking for characteristics within each methodology that, as per best practices, lead to effective assessment.

1. How were the reviewed cyber risk assessment methodologies evaluated?
2. What is the potential effectiveness of the studied cyber risk assessment methodologies?

### 3.2. Literature Search Process

A three-step search process was followed to search for scholarly contributions on cyber risk assessment for CPS. The first search step followed the criteria specified in Table 1. The search keywords were identified through brainstorming and preliminary research. They were expanded with synonyms and refined by testing them on database search queries. The keywords in the search string were searched in the full text of each research paper to avoid missing relevant publications. This initial search resulted in a total of 1329 publications.

Table 1. Search Sources

| Sources of Research Paper | Research Database | Initial Paper Count |
|---|---|---|
| | IEEE Xplore | 397: articles (106), conference papers (291) |
| | IEICE | 30: articles (30) |
| | Science Direct | 358: articles (358) |
| | SpringerLink | 408: articles (146), conference papers (262) |
| | MDPI | 57: articles (57) |
| | ACM | 79: articles (16), conference papers (63) |
| Search string keywords | ("cybersecurity risk" OR "cyber risk" OR "cyber-to-physical risk") AND ("assessment" OR "analysis" OR "evaluation") AND ("cyber-physical system" OR "critical infrastructure" OR "SCADA" OR "DSC" OR "industrial control system") | |
| Search items | Journals' articles, conference papers | |
| Search applied on | Full text | |
| Language | English | |
| Publication period | 2014-2023 October | |
| Initial paper count | 1329 | |
| 2nd search paper count | (328): 101 IEEE Xplore, 6 IEICE, 102 Science Direct, 56 SpringerLink, 28 MDPI, 35 ACM | |
| 3rd search paper count | (275): 95 IEEE Xplore, 3 IEICE, 92 Science Direct, 48 SpringerLink, 17 MDPI, 20 ACM | |
| Papers selected for review | (28): 9 IEEE Xplore, 1 IEICE, 7 Science Direct, 5 SpringerLink, 4 MDPI, 2 ACM | |

In the following search step, the abstract of each paper gathered in the initial search was scrutinized based on the inclusion and exclusion criteria presented in Table 2. At the end of this, 1001 of the 1329 papers failed to meet the defined inclusion criteria; hence, they were excluded. In the third search step, some sections in the remaining 328 papers were examined: the sections explaining the proposed approach components and steps and results and evaluation were examined against the level of contribution criteria, depicted in Table 2. After applying this criterion, 53 studies were filtered out. The 275 papers left were preferenced based on their contribution level. From this preferencing, twenty-eight publications were selected for the review: the top fourteen papers (50%) published in recent years, 2022 and 2023, and the top fourteen (50%) published between 2014 and 2021. This approach was taken to balance the inclusion of the current with foundational research. The papers selected for review are listed in the following section.

Table 2. Inclusion and exclusion criteria

| Inclusion Criteria | Exclusion Criteria |
|---|---|
| Second search step ||
| The central theme: papers that focus primarily on cyber risk assessment | • Papers that examine cybersecurity from more than one aspect, including risk assessment, but the discussion is not dedicated to the risk assessment process.<br>• Papers addressing challenges in the risk management process, but the discussion of the risk assessment phase is limited.<br>• Papers that are not specific about the category of information security on which the risks are assessed. |
| Relevance: the content is relevant to risk assessment related to the cybersecurity domain in the context of CPS, explicitly concentrating on cyber risks arising from external attacks. | • Articles that address risk assessment on dimensions other than security, such as resilience, privacy, and reliability.<br>• Articles dedicated to risk assessment in security domains other than cybersecurity, such as physical, operational, and personal security.<br>• Articles on assessing cyber risks arising from insider attacks.<br>• Articles that assess the cyber risks of a CPS system under the design phase of the system life cycle. |
| Third search step ||
| Level of contribution: papers that add a unique, non-replicated, and sufficiently detailed assessment approach. | • Risk assessment approaches replicating guidelines found in risk assessment standards codified by regulatory bodies.<br>• Papers that do not provide sufficient details about their risk assessment methodology component and steps. |

## 4. RESULTS AND DISCUSSION

This section discusses the review findings of the selected twenty-eight papers, which are listed in Table 3. It answers the two research questions mentioned in 3.1. The first sub-section (4.1) answers how the reviewed risk assessment approaches are evaluated. The second sub-section (4.2) synthesizes results on the extent to which the studied risk methodologies meet the requirements of effective risk assessment.

Table 3. The cyber risk assessment approaches selected for review

| S. No | Study title & Reference | Publication Type | Publication Year |
|---|---|---|---|
| 1 | Security-Oriented Cyber-Physical Risk Assessment for Cyberattacks on Distribution System [32] | Journal article | 2023 |
| 2 | Risk Assessments in Virtual Power Plants with NESCOR Criteria, Practical Application, Advantages and Disadvantages [33] | Journal article | 2023 |
| 3 | Assessing Cyber Risk in Cyber-Physical Systems Using the ATT&CK Framework [34] | Journal article | 2023 |
| 4 | Dependency-based security risk assessment for cyber-physical systems [35] | Journal article | 2023 |
| 5 | Development of the framework for quantitative cyber risk assessment in nuclear facilities [36] | Journal article | 2023 |
| 6 | Exploring the Cyber-Physical Threat Landscape of Water Systems: A Socio-Technical Modelling Approach [37] | Journal article | 2023 |
| 7 | A Study of The Risk Quantification Method of Cyber-Physical Systems focusing on Direct-Access Attacks to In-vehicle networks [38] | Journal article | 2023 |
| 8 | A Quantitative Risk Assessment Model for Distribution Cyber Physical System under Cyber Attack [39] | Journal article | 2023 |
| 9 | Cyber security risk assessment in autonomous shipping [40] | Journal article | 2022 |
| 10 | Threat modelling for industrial cyber physical systems in the era of smart manufacturing [41] | Journal article | 2022 |
| 11 | An Integrated cyber security risk management framework and risk predication for the critical infrastructure protection [42] | Journal article | 2022 |
| 12 | A Cyber-Physical Risk Assessment Approach for Internet of Things Enabled Transportation Infrastructure [43] | Journal article | 2022 |
| 13 | Managing cybersecurity risks of cyber-physical systems: The MARISMA CPS pattern [44] | Journal article | 2022 |
| 14 | Quantitative Risk Assessment of Cyber Attacks on Cyber-Physical Systems using Attack Graphs [45] | Conference Paper | 2022 |
| 15 | Qualitative Risk Assessment of Cybersecurity and Development of Vulnerability Enhancement Plans in consideration of digitalized ship [46] | Journal article | 2021 |
| 16 | Harmonizing safety and security risk analysis and prevention in cyber-physical systems [47] | Journal article | 2021 |
| 17 | Cyber security risk assessment for seaports: A case study of a container port [48] | Journal article | 2021 |
| 18 | Model-based risk assessment for cyber physical systems security [49] | Journal article | 2020 |
| 19 | Bayesian Network Based C2P Risk Assessment for Cyber-Physical Systems [50] | Journal article | 2020 |
| 20 | Quantitative Risk Modelling and Analysis for Large-Scale Cyber-Physical Systems [51] | Conference Paper | 2020 |
| 21 | MaCRA: a model-based framework for maritime cyber-risk assessment [52] | Journal article | 2019 |
| 22 | Risk Assessment for Cyber Security of Manufacturing Systems: A Game Theory Approach [53] | Journal article | 2019 |
| 23 | Risk Assessment for Cyber Attacks in Feeder Automation System [54] | Conference Paper | 2018 |
| 24 | Security risk assessment framework for smart car using the attack tree analysis [55] | Journal article | 2018 |
| 25 | Network Topology Risk Assessment of Stealthy Cyber Attacks on Advanced Metering Infrastructure Networks [56] | Conference Paper | 2017 |
| 26 | Attack-Defence Trees based cyber security analysis for CPS [57] | Conference Paper | 2016 |
| 27 | Risk assessment framework for power control systems with PMU-based intrusion response system [58] | Journal article | 2015 |
| 28 | Cyber-related Risk Assessment and Critical Asset Identification in Power Grids [59] | Conference Paper | 2014 |

## 4.1. Evaluation Methods of the Studied Cyber Risk Assessment Approaches

The reviewed cyber risk assessment approaches were demonstrated either through theoretical case studies of a CPS system subjected to cyber-attack scenarios or by applying them to real-world CPS systems. As in Figure 1, 89% of the risk assessment methodologies were demonstrated through case studies, and 11% were applied to real-world CPS systems. The case studies were informed by different sources of cyber-attacks. For example, fifteen case studies were inspired by practical demonstrations of cyber-attacks on a CPS system through simulations or experimental testbeds. The remaining ten were not based on any empirical setting but were informed by the authors' knowledge of CPS system security.

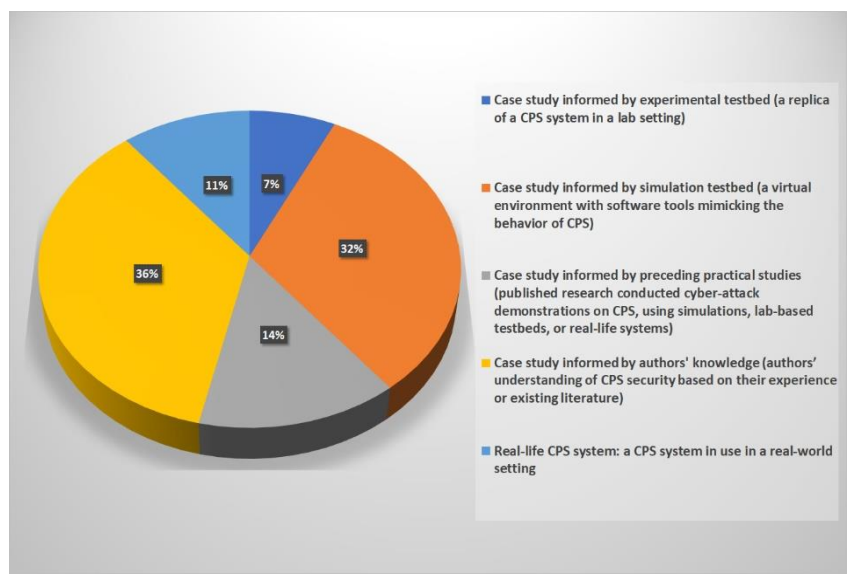

Figure 1. Demonstration methods and settings of the reviewed risk assessment methodologies

While informed by realistic sources of cyber-attacks encountered in CPS, case studies nonetheless do not reflect the actual complexity of a real-world CPS. For example, the case studies drawn from experimental and simulated testbeds are narrowed to a single or few most representative attack scenarios, often failing to reflect the constantly changing cyber threat landscape. Additionally, the case studies developed from the authors' knowledge might overlook the practical aspects of the CPS system and are susceptible to the authors' subjective interpretations and heuristics. Given these limitations, the thoroughness of the evaluation process based on case studies requires further examination.

In the three risk assessment approaches demonstrated in real-world settings [42,44,48], the CPS systems were accessed only once for the proactive risk assessment. They were not revisited for feedback on the effectiveness of the risk assessment method after the identified risks materialized. For a rigorous evaluation of a risk assessment methodology, comparing the proactive risk assessment results to the reactive assessment conducted after the identified risks materialize is required. Revaluation validates whether the proactive risk assessment effectively identified and assessed the risks encountered. Additionally, independent security experts or system owners' analyses is needed to evaluate further and validate the risk assessment methodologies.

As explained in Section 1, multiple criteria indicate an effective risk assessment, including validity (determined by the degree of risk assessment results accuracy and handling uncertainties), reliability (determined by repeatability and reproducibility). Compared to these criteria, no methodology among the analyzed papers was found to be thoroughly evaluated for effectiveness. The evaluations in the reviewed studies predominantly focused on the applicability of the proposed approaches by showcasing their capabilities in performing the intended functionalities. The studied risk assessment approaches were not evaluated in terms of pivotal criteria that directly impact risk management decisions, like the validity and reliability. Validity and reliability indicate how correctly and consistently the risk assessment method understand and express risk [20]. The validity of the risk assessment in producing risk assessment results with an acceptable extent of accuracy plays a significant role in better prioritizing and managing the most significant risks. Inaccurate assessment creates a flawed picture of the organization's security posture, leading to inappropriate risk management decisions [60]. A subsequent result of inappropriate risk management decisions is a reduced organization's capability to respond to cyber risks; prevent them, mitigate their impact, or effectively handle them when they materialize.

Given the limitations in how the studied cyber risk assessment approaches were evaluated for effectiveness, demonstrating cyber risk assessment methodologies in a live CPS is needed for gaining confidence in the method's applicability to real-world settings. Equally important is engaging independent experts or system owners in evaluating the risk assessment method. Their feedback enables a realistic evaluation that reflects the system's intricacies and practical challenges and isolates biases. The effectiveness of a risk assessment methodology should be thoroughly evaluated across multiple criteria.

## 4.2. The Potential Effectiveness of the Studied Cyber Risk Assessment Methodologies

As discussed in Section 4.1, validity and reliability are pivotal criteria in determining the effectiveness of risk assessment. These criteria influence decision-makers' confidence in the risk assessment results when making risk management decisions. The literature has limited attention to the sub-criteria or the requirements for validity and reliability of the risk assessment. In the studies where these criteria are mentioned, there is limited specification about the features of the risk assessment method that indicate reliable and valid assessments. The review study [61] considered the source of data the risk assessment techniques rely on to rate their reliability, specifically repeatability, for the nuclear industry. The NIST guide for risk assessment mentioned that the age of information used in assessing risk is a particular concern when evaluating the validity of assessment results [17]. NIST also stated that the precision of the risk assessment is affected by the subjectivity, trustworthiness of the data drawn upon, and the interpretation of assessment results. Cherdantseva [62] described the essence of risk assessment methods in terms of aim, risk factors measurement, sources of data, and tool support.

Based on these theoretical views, for a risk assessment to be an effective, it is required to meet these key requirements: well-definition of the risk model, trustworthiness of the cyber threat intelligence on which it relies, consistency of cyber threat intelligence interpretations, and automation support. This following sub-sections investigate the potential effectiveness of the studied methodologies by examining how they align with these key factors.

### 4.2.1. Well-Definition of the Risk Model

The Risk model is a component of the risk assessment methodology that describes the risk determination process. It defines assessable risk factors and their relationships in a formula or a matrix. Risk factors are inputs to determining the overall risk value or level in risk assessments. Typical risk factors include threat, vulnerability, impact, and likelihood. Risk factors are usually decomposed into characteristics to gauge their values (e.g., impact is decomposed into impact on cyber and physical systems). Well-definition of the risk model prior to conducting risk assessments could have a significant effect on the effectiveness of the assessment results.

The risk models in the reviewed risk assessment methodologies revealed limitations that could impact the validity and reliability of the overall risk determination. First, using non-inclusive attributes to gauge the risk factors like impact and likelihood, the prevalent metrics in the literature, leads to imprecise assessment. For instance, some approaches assessed the impact of potential cyber-attack risk on either a cyber or a physical system layer but not both [46,51,57,58]. Similarly, in some approaches, the likelihood of a potential cyber-attack occurring was interpreted as the likelihood of successfully exploiting a vulnerability [57,58]. In a real attack scenario, successful vulnerability exploitation does not always result in a successful attack or impact on the system. Another gap is the determination of the overall risk based on only one factor, which could result in underestimating the cyber threats, like in [47,51,52,56]. Some approaches might produce overestimated risk values because they do not take into account the role of the defender's countermeasures in mitigating the system's exposure to risks [45,58].

### 4.2.2. Trustworthiness of Cyber Threat Intelligence

The trustworthiness of cyber threat information on which the risk assessment relies for identifying threats and gauging risk factors values affects the validity of the assessment results. Up-to-date and relevant threat intelligence contributes to correctly representing the cyber risks the organization is exposed to.

In 50% of the approaches studied, risk assessment was based on a single source of cyber threat information, as presented in Figure 2. Specifically, risk determination in six methodologies drew upon the demonstration of cyber-attack scenarios. The assessment in three methodologies utilized public security databases or reports. In five approaches, the assessment was informed by the judgment of security experts, the system owners, or the authors' estimation. The other 50% of the surveyed approaches combined multiple sources to identify and estimate cyber risks. For instance, real-world CPS system historical cyber-attack information appeared in three approaches, complemented by other sources such as public security databases or the judgments of security experts.

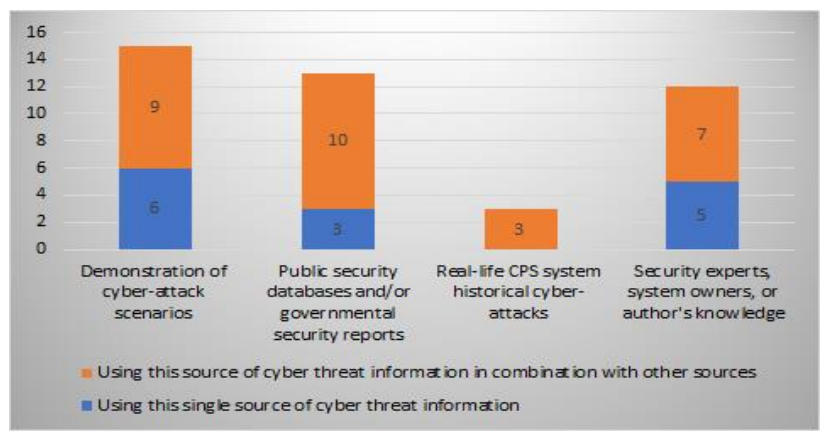

Figure 2. Cyber threat intelligence sources on which the studied risk assessment methodologies relied

Although these diverse cyber threat intelligence sources form a broader view of the cyber threat landscape, they possess limitations that diminish the assessment capacity to assess potential cyber risks correctly, as shown in Table 4. For instance, the cyber-threat information published in public security databases or governmental reports does not reflect the organization's unique experience. Public cyber threat intelligence can become outdated due to the evolving threat landscape, leading to an inaccurate representation of the risks to which the organization is exposed. Given the significant underreporting of OT incidents and a lack of scientifically verified data on cyber-physical attacks [43], such datasets are unreliable.

A demonstration of cyber-attack scenarios in a simulated or experimental testbed does not reflect the complexity of a real-world CPS system and the risks to which it is exposed. It is challenging for the threat event parameters in the attack scenario to keep up with the constantly changing threat landscape [37]. In the case of insufficient knowledge about threat characteristics, the authors either omitted them in their attack scenarios or relied on subjective assumptions. The scenario approach is further narrowed when researchers describe potential threats into a single or few most representative scenarios.

The past cyber-attacks experienced by a real-world CPS system do not correctly assess a potential cyber risk because the factors that led to attacks in the past are changing. The historical cyberattacks-based approach lacks real-time learning of a changing threat landscape. Furthermore, the assessment based on querying security experts, system owners, or researchers' judgment is susceptible to biases and misleading heuristics. The bias arises if the judgment is based on personal experiences or beliefs and what comes to mind rather than objective and quantifiable data.

Table 4. Limitations in the cyber threat intelligence

| Limitations in the cyber threat intelligence trustworthiness | Assessment approaches relied on this threat intelligence source (References) |
| --- | --- |
| The public threat intelligence does not reflect the organization's unique experience and can become outdated. | [33,34,38,39,41,42,45,47,50–54] |
| Scenario-based approach does not accurately parameterize the threat events or represent the actual complexity of a CPS system. | [32,37-39,41,43,45,47,49,50,54–58] |
| Past cyber-attacks experienced by a real-world CPS system are nonrepresentative and lack real-time learning of the threat landscape. | [42,44,48] |
| Security experts, system owners, or researchers' judgments are susceptible to biases and misleading heuristics. | [32–36,40,41,44,46,48,58,59] |

Based on the above analysis, the cyber threat intelligence sources on which the studied risk assessment approaches rely provide incomplete knowledge of the potential cyber risks. These cyber threat information sources can be outdated or irrelevant, leading to incorrect or risk assessment results with high uncertainty.

**4.2.3. Consistency of Cyber Threat Intelligence Interpretations**

The validity and reliability of the risk assessment can be further affected by how consistently the cyber threat intelligence is interpreted across different assessors. Inconsistent interpretation of cyber threat information can lead to discrepancies and high level of uncertainty. Inconsistent interpretation arises from individual assessors' bias, where their beliefs influence how they assign meaning to the information of a specific cyber-threat. Another source of inconsistency is the ambiguity in the cyber threat information, which may lead to different analysts coming to different conclusions. Furthermore, when a threat intelligence source provides incomplete information about a specific threat, assessors make assumptions, which might be subjective and varied, to fill these gaps.

Although the reviewed approaches provided detailed steps to guide conducting the risk assessment, this is inadequate to address the probable inconsistencies in cyber threat intelligence interpretations. These approaches lacked methods and techniques for structuring how to interpret lessons about potential cyber risks, leaving room for varied interpretations. Given that 50% of the surveyed approaches draw upon two or more cyber threat intelligence sources, the potential for variability in interpretations is heightened. The lack of standardization methods affects the uniformity of the assessment and, hence, its validity and reliability.

### 4.2.4. Automation Support

A further constraint to achieving reliable risk assessment and valid results is the manual efforts required to conduct the assessment process, especially if it is not supported by a software tool. Manual risk assessment methodologies are less scalable and prone to human errors and inconsistencies. Supporting the risk assessment methodology with automation that facilitates the risk assessment steps, from data acquisition to risk calculation and dissemination, is crucial for reliable assessment, especially in a highly interconnected CPS infrastructure.

Among the risk assessment methodologies reviewed in this study, 75% lack software support, relying entirely on manual efforts. In the remaining seven approaches [34,40–42,44,45,56], the authors developed software tools or created Python or MATLAB scripts to facilitate the risk assessment process. However, the intelligence these tools and scripts provide is limited. For instance, in identifying the potential cyber risks, these tools do not complement the assessors' knowledge with real-time learning from the publicly available threat intelligence sources and the cybersecurity incidents the organization undergoes. Similarly, in risk analysis, they do not automate the modeling and simulation of threat scenarios based on what they learn.

The automation support in the reviewed approaches has not yet reached a level of maturity where it facilitates a risk assessment process, from preparation for the risk assessment to conducting it, communicating its results, and maintaining it over time. With this lack of automation support, human assessors cannot effectively analyze the relevant cyber threat intelligence sources in real time. As a result, the ability of the studied assessment methodologies to maintain consistency and adapt to the constantly changing threat landscape will be hindered, thereby limiting assessment reliability and validity.

### 4.2.5. The Potential Effectiveness of the Risk Assessment Approaches

In light of the preceding analysis, the reviewed cyber risk assessment methodologies revealed limited effectiveness in assessing cyber risks. They fall short in conforming to different factors crucial for ensuring assessment effectiveness. Some methodologies lack well-defined risk models to evaluate the risk factors or the overall risk. Another shortcoming relates to the trustworthiness of cyber threat intelligence sources they rely on, which can be outdated, nonrepresentative, or imprecise. Another effectiveness requirement the studied risk assessment methodologies failed to meet is addressing the probable inconsistencies in cyber threat intelligence interpretations. Finally, most reviewed risk assessment methodologies rely solely on manual efforts.

## 5. REAL-TIME LEARNING FROM CYBERSECURITY INCIDENTS FOR IMPROVED CYBER RISK ASSESSMENT EFFECTIVENESS

The more challenging factors impacting the risk assessment effectiveness, which have not been adequately discussed in the existing literature, are (1) the nature of cyber threat information sources on which the risk assessment relies and (2) the way human analysts interpret this information. The trustworthiness of cyber threat information contributes greatly to risk assessment effectiveness and must be up-to-date, relevant, representative, and precise. Consistency in interpreting the cyber threat information across different assessors is also crucial for effective assessment. Minimizing inconsistencies in the assessors' interpretations is challenging but not impossible if the sources of these inconsistencies are identified.

The multiple requirements for trustworthy cyber threat information and consistent interpretations underscore the need for a cyber risk assessment approach of more than a component collaboration to improve the risk assessment's effectiveness. Such an approach could leverage cybersecurity incidents the organization undergoes as a primary source to learn about potential cyber risks. As discussed in Section 2, real-time learning from cybersecurity incidents can complement the assessors' knowledge with up-to-date information about cyber risks representing an organization's unique experience and the complexities of real-world CPS systems. Automation support is crucial to enabling real-time learning from cybersecurity incidents and other cyber threat intelligence sources. A machine learning-based software tool is recommended to enable learning from the cyber incident as it occurs and update the risk assessment process. As discussed in Section 2, machine learning can reduce manual efforts, thereby better structuring the acquisition, retainment, dissemination, and application of knowledge about cyber risks. ML capabilities contribute to minimizing human errors and inconsistencies in cyber-threat interpretations.

## 6. CONCLUSION

This literature review examined how the studied cyber risk assessment approaches were evaluated. It also investigated to what extent these approaches meet the requirements of effective risk assessment. The findings revealed that the effectiveness of the examined risk assessment approaches was predominantly evaluated in terms of applicability but not other pivotal criteria like validity and reliability. The existing cyber risk assessment methodologies have been found to fall short in conforming to the assessment effectiveness requirements, which include a well-defined risk model, trustworthy cyber threat intelligence, consistent cyber threat intelligence interpretations, and automation support. This study identified the untrustworthiness of cyber threat intelligence and the inconsistent interpretations of this cyber threat information as the most challenging gaps that limit risk assessment effectiveness. It proposed real-time learning from cybersecurity incidents to overcome these limitations. Future research will deeply investigate the contribution of the given recommendations in addressing the most challenging gaps impacting the assessment effectiveness.

**ACKNOWLEDGEMENTS**

The authors would like to thank Sultan Qaboos University for their support throughout the various phases of this research. The resources provided by the university facilitated the completion and publication of this research paper.